\documentclass[11pt]{article}
\usepackage{amssymb}
\usepackage{amsmath}
\usepackage{amscd}
\usepackage{latexsym}

\catcode`@=11
\renewcommand{\section}{\@startsection{section}{1}{0pt}{\medskipamount}
{\medskipamount}{\large\bf}}
\numberwithin{equation}{section}
\catcode`@=12

\def\de{\delta}
\def\ve{\varepsilon}
\def\th{\theta}

\def\o{\omega}

\newcommand{\R}{\mathbb R}
\newcommand{\Acal}{{\cal A}}
\newcommand{\Fcal}{{\cal F}}

\newcommand{\Fct}{{\widetilde{\cal F}}}
\newcommand{\Act}{{\widetilde{\cal A}}}
\newcommand{\Et}{{\widetilde{E}}}
\newcommand{\Bt}{{\widetilde{B}}}

\def\N2{$N{=}2$}
\def\pa{\mbox{$\partial$}}
\def\diff{\mbox{d}}
\def\tr{{\rm tr}}
\def\sfrac#1#2{{\textstyle\frac{#1}{#2}}}
\def\>{\rangle}
\def\<{\langle}
\def\+{\dagger}
\def\={\ =\ }

\begin{document}

\begin{titlepage}
\setcounter{page}{0}

\phantom{.}
\vskip 1.5cm

\begin{center}

{\LARGE{\bf Solutions to Yang-Mills equations 
\\[12pt]
on four-dimensional de Sitter space}}

\vspace{20mm}

{\Large
Tatiana A. Ivanova${}^*$, \ Olaf Lechtenfeld${}^{\+\times}$ \ and \  Alexander D. Popov${}^\+$
}\\[10mm]
\noindent ${}^*${\em
Bogoliubov Laboratory of Theoretical Physics, JINR\\
141980 Dubna, Moscow Region, Russia
}\\
{Email: ita@theor.jinr.ru
}\\[5mm]
\noindent ${}^\+${\em
Institut f\"ur Theoretische Physik, 
Leibniz Universit\"at Hannover \\
Appelstra\ss{}e 2, 30167 Hannover, Germany
}\\
{Email: alexander.popov@itp.uni-hannover.de}
\\[5mm]
\noindent ${}^\times${\em
Riemann Center for Geometry and Physics,
Leibniz Universit\"at Hannover \\
Appelstra\ss{}e 2, 30167 Hannover, Germany
}\\
{Email: olaf.lechtenfeld@itp.uni-hannover.de}

\vspace{20mm}

\begin{abstract}
\noindent 
We consider pure SU(2) Yang--Mills theory on four-dimensional de Sitter space dS$_4$ 
and construct a smooth and spatially homogeneous magnetic solution to the Yang--Mills equations.
Slicing dS$_4$ as $\R\times S^3$, via an SU(2)-equivariant ansatz we reduce the Yang--Mills equations 
to ordinary matrix differential equations and further to Newtonian dynamics in a double-well potential. 
Its local maximum yields a Yang--Mills solution whose color-magnetic field at time $\tau\in\R$ 
is given by $\widetilde{B}_a=-\frac12 I_a/(R^2\cosh^2\!\tau)$, where $I_a$ for $a=1,2,3$ are 
the SU(2) generators and $R$ is the de Sitter radius. At any moment, this spatially homogeneous configuration 
has finite energy, but its action is also finite and of the value $-\frac12j(j{+}1)(2j{+}1)\pi^3$ in a 
spin-$j$ representation. Similarly, the double-well bounce produces a family of homogeneous finite-action 
electric-magnetic solutions with the same energy. There is a continuum of other solutions whose energy and 
action extend down to zero.
\end{abstract}

\vspace{12mm}


\end{center}
\end{titlepage}

\section{Introduction and summary}

\noindent 
Yang--Mills theory with Higgs fields governs three fundamental forces of Nature. It has a number
of particle-like solutions such as vortices, magnetic monopoles and instantons~\cite{1, 2, 3}. 
In Minkowski space $\R^{3,1}$ smooth vortex and monopole solutions can be constructed only in the presence of Higgs fields.
Magnetic monopoles play a key role in the dual superconductor mechanism for the confinement of quarks in QCD
based on the condensation of non-Abelian monopoles~\cite{4}. However, there are no Higgs fields in QCD, and without them 
all monopole solutions of pure Yang--Mills theory on $\R^{3,1}$ are singular Abelian monopoles, with the U(1)
gauge group embedded into a higher-dimensional non-Abelian gauge group, e.g.\ SU(3) for the QCD case. Furthermore,
a number of theorems rule out any static, real, finite-energy solution of pure SU(2) Yang--Mills theory on $\R^{3,1}$~\cite{5,6}.
Under some mild assumptions the non-existence can be extended also to time-dependent finite-energy solutions in pure Yang--Mills theory 
on $\R^{3,1}$~\cite{7, 8}.

The still elusive quantitative understanding of the confinement mechanism in Minkowski space is not the only problem. 
In most inflationary models the early and late expansion of the universe is approximated by a dS$_4$ phase. 
Therefore, it is important to understand de Sitter vacua of supergravity and string theory~\cite{KKLT}. 
Such vacua include one or more anti-D3-branes~\cite{KKLMMT} (for recent results see, e.g.,~\cite{BDKVW}).
In fact, gauge theories in dS$_4$ occur naturally in string-theoretic constructions from stacks of branes or from compactifications. 
Hence, constructing explicit solutions and studying their physical effects in gauge theories on de Sitter space are also important 
for understanding the early universe and its evolution.

So on the one hand, Minkowski space does not seem to admit physical non-Abelian field configurations.
On the other hand, our universe appears to be asymptotically de Sitter (not Minkowski) at very early and very large times.
This is a strong argument for searching finite-energy solutions in pure Yang--Mills theory on four-dimensional de Sitter space dS$_4$. 
The construction of such solutions is the goal of our paper.\footnote{
Note that we consider the spacetime background as non-dynamical, i.e.\ we ignore the backreaction on it.
The coupled system is governed by the Einstein--Yang--Mills equations
(for numerical solutions see e.g.\ the review~\cite{9} and references therein).
However, in this more general setup it is practically impossible to obtain analytic solutions.}

We will not only show that smooth non-Abelian solutions with finite energy indeed exist in pure Yang--Mills theory on dS$_4$,
but we will also construct them analytically in a rather simple geometric form.
On the spatial $S^3$ slices of dS$_4$, the gauge potential as well as the color-electric and -magnetic fields are constant,
analogous to the Dirac monopole on $\R^3$ restricted to $S^2$ or the Yang monopole on $\R^5$ restricted to $S^4$~\cite{Yang}.
Furthermore, their temporal variation is such that the action functional is also finite.
One may hope that such configurations will help to get a quantitative understanding of QCD confinement in a de Sitter background.
\bigskip

\section{Description of de Sitter space dS$_4$}

\noindent 
Topologically, de Sitter space is $\R\times S^3$, and it can be embedded into 5-dimensional
Minkowski space $\R^{4,1}$ by the help of
\begin{equation}\label{2.1}
\de_{ij}y^iy^j - (y^5)^2 \= R^2 \qquad\textrm{where}\quad i,j=1,\ldots,4\ .
\end{equation}
One can parametrize dS$_4$ with global coordinates $(\tau , \th^a)$,
$a=1, 2, 3$, by setting (see e.g.~\cite{10})
\begin{equation}\label{2.2}
y^i= R\,\o^i\cosh\tau\ ,\quad y^5=R\sinh\tau \qquad\textrm{with}\quad\tau\in\R
\end{equation}
for
\begin{equation}\label{2.3}
\o^1=\cos\th_1\ ,\ \ \o^2=\sin\th_1\cos\th_2\ ,\ \ 
\o^3=\sin\th_1\sin\th_2\cos\th_3\ ,\ \ \o^4=\sin\th_1\sin\th_2\sin\th_3\ ,
\end{equation}
where $0\le\th_1, \th_2<\pi$ and $0\le\th_3\le 2\pi$. Then from the flat metric
on $\R^{4,1}$ one obtains the induced  metric on dS$_4$,
\begin{equation}\label{2.4}
\diff s^2 \= R^2\,\bigl( - \diff\tau^2 + \cosh^2\!\tau\,\de_{ab}e^a e^b \bigr)
\= -\diff\tilde\tau^2 + R^2\cosh^2\!\sfrac{\tilde\tau}{R}\,\diff\Omega^2_3
\qquad\textrm{with}\quad \tilde\tau = R\,\tau \ ,
\end{equation}
where $\diff\Omega^2_3$ is the metric on the unit sphere $S^3\cong\,$SU(2) and $\{e^a\}$ is an
orthonormal basis of left-invariant one-forms on $S^3$ satisfying
\begin{equation}\label{2.5}
\diff e^a + \ve^a_{bc}\,e^b\wedge e^c\=0\ .
\end{equation}

We rewrite the metric (\ref{2.4}) in conformal coordinates $(t,\th^a)$ 
by the time reparametrization~\cite{10}
\begin{equation}\label{2.6}
t\=\arctan (\sinh\tau)\=2\arctan(\tanh\sfrac{\tau}{2})
\qquad\Longleftrightarrow\qquad 
\frac{\diff\tau}{\diff t} \= \cosh\tau \= \frac{1}{\cos t}\ ,
\end{equation}
in which $\tau\in (-\infty, \infty )$ corresponds to $t\in (-\sfrac{\pi}{2}, \sfrac{\pi}{2})$. 
The metric (\ref{2.4}) in these coordinates reads
\begin{equation}\label{2.7}
\diff s^2 \= \frac{R^2}{\cos^2\!t}\,\bigl(- \diff t^2 + \de_{ab}e^ae^b \bigr)
\= \frac{R^2}{\cos^2\!t}\,\diff s^2_{\textrm{cyl}}  \ ,
\end{equation}
where
\begin{equation}\label{2.8}
\diff s^2_{\textrm{cyl}} \= - \diff t^2 + \de_{ab}e^ae^b
\end{equation}
is the standard metric on the Lorentzian cylinder $\R\times S^3$. 
Hence, four-dimensional de Sitter space is conformally equivalent to the finite cylinder 
${\cal I}\times S^3$ with the metric (\ref{2.8}),
where ${\cal I}$ is the interval $(-\frac{\pi}{2}, \frac{\pi}{2})$ parametrized by~$t$.

\bigskip

\section{Reduction of Yang--Mills to matrix equations and to double-well dynamics}

\noindent 
Since the Yang--Mills equations are conformally invariant, 
their solutions on de Sitter space can be obtained by solving
the equations on ${\cal I}\times S^3$ with the cylindrical metric (\ref{2.8}).
Therefore, we will consider rank-$N$ vector bundles over this cylinder
${\cal I}\times S^3$ with the de Sitter (\ref{2.7}) or cylindrical (\ref{2.8}) metric. 
Our gauge potentials $\Acal$ and the gauge fields $\Fcal=\diff\Acal + \Acal\wedge\Acal$ 
take values in the Lie algebra $su(N)$. The conformal boundary of dS$_4$
consists of the two 3-spheres at $t=\pm\frac{\pi}{2}$ or, equivalently, at $\tau=\pm\infty$. 
On manifolds $M$ with a nonempty boundary $\pa M$, the group of gauge transformations 
is naturally restricted to the identity when reaching $\pa M$ (see e.g.~\cite{11}).
This corresponds to a framing of the gauge bundle over the boundary.
For our case, this means allowing only gauge-group elements $g(y)$ obeying 
$g(\pa M)=\,$Id on $\pa M=S^3_{t=\pm\frac{\pi}{2}}=S^3_{\tau=\pm\infty}$.

In order to obtain explicit solutions we use the SU(2)-equivariant ansatz (cf.~\cite{12,13,14})
\begin{equation}\label{3.1}
\Acal(y) \= X_a(t)\,e^a
\end{equation}
for the $su(N)$-valued gauge potential $\Acal$ in the temporal gauge
$\Acal_0\equiv\Acal_t=0=\Acal_\tau$. Here, $X_a(t)$ are three
$su(N)$-valued matrices depending only on $t\in{\cal I}$, and $e^a$
are basis one-forms on $S^3$ satisfying (\ref{2.5}). 
The corresponding gauge field reads
\begin{equation}\label{3.2}
\Fcal \= \Fcal_{0a}\,e^0\wedge e^a + \sfrac12 \Fcal_{bc}\,e^b\wedge e^c
\= \dot{X}_a\,e^0\wedge e^a + \sfrac12\bigl(-2\ve^a_{bc}X_a + [X_b, X_c]\bigr)e^b\wedge e^c\ ,
\end{equation}
where $\dot{X}_a:=\diff{X}_a/\diff t$ and $e^0:=\diff t$. 
It is not difficult to show (see e.g.~\cite{14}) that the Yang--Mills equations on ${\cal I}\times S^3$ 
after substituting (\ref{3.1}) and (\ref{3.2}) reduce to the ordinary matrix differential equations
\begin{equation}\label{3.3}
\ddot X_a \= -4 X_a+ 3 \ve_{abc}\, [X_b, X_c] - \bigl [X_b, [X_a,X_b]\bigr]\ .
\end{equation}

To be more concrete, we let the gauge potential and fields take values in an $su(2)$ subalgebra
of $su(N)$. In other words, we pick three SU(2) generators $I_a$ in a spin-$j$ representation
embedded into $su(N)$ obeying
\begin{equation}\label{3.4}
[I_b, I_c] \= 2\,\ve^a_{bc}I_a \qquad\textrm{and}\qquad \tr(I_aI_b)\=-4\,C(j)\,\de_{ab}
\qquad\textrm{where}\qquad C(j)=\sfrac13j(j{+}1)(2j{+}1)
\end{equation}
is the second-order Dynkin index of the representation.
Explicit solutions to the matrix equations (\ref{3.3}) can then be found with the natural choice
\begin{equation}\label{3.5}
X_a(t) \= \sfrac12\, \bigl(1+\psi(t)\bigr)\,I_a\ ,
\end{equation}
where $\psi(t)$ is a real function. The Yang--Mills equations finally boil down to
\begin{equation}\label{3.7}
\ddot\psi \= 2\,\psi\,(1+\psi)(1-\psi) \= -\sfrac{\diff V}{\diff\psi}
\qquad\textrm{for}\quad V(\psi)\=\sfrac12(1-\psi^2)^2\ .
\end{equation}
This is Newton's equation for a particle in a double-well potential, whose solutions are well known.

The simplest ones are constant at the critical points of~$V$, i.e.
\begin{equation} \label{psiconst}
\psi(t) = \pm1 \ \textrm{(minima, $V{=}0$)} \qquad\textrm{and}\qquad 
\psi(t) = 0 \ \textrm{(local maximum, $V{=}\sfrac12$)}\ .
\end{equation}
A prominent nontrivial solution is the bounce,
\begin{equation}\label{3.9}
\psi(t)\= \sqrt{2}\,\textrm{sech}\bigl(\sqrt{2}(t{-}t_0)\bigr) 
\= \frac{\sqrt{2}}{\cosh\bigl(\sqrt{2}(t{-}t_0)\bigr)}\ ,
\end{equation}
which makes an excursion from $(\psi{=}0,V{=}0)$ at $t{=}{-}\infty$ 
to $(\psi{=}\sqrt{2},V{=}0)$ at $t{=}t_0$ and back at $t{=}\infty$.
An anti-bounce is given by $-\psi(t)$.
In addition, there is a continuum of periodic solutions oscillating either about $\psi=\pm1$
or exploring both wells, which are given by Jacobi elliptic functions.
Usually, the moduli parameter~$t_0$ is trivial because of time translation invariance in~(\ref{3.7}).
However, since for de Sitter space according to (\ref{2.6}) we consider the solutions $\psi(t)$ 
only in the interval ${\cal I}=(-\frac{\pi}{2},\frac{\pi}{2})$ without imposing boundary conditions, 
the value of $t_0\in\R$ makes a difference. It allows us to pick a segment of length~$\pi$ anywhere 
on the profile of the bounce, not necessarily including its minimum.
Finally, we remark that the Newtonian energy conservation produces the relation
\begin{equation}\label{3.10}
\sfrac12\,\dot\psi^2 \= V_0-V(\psi) \= V_0-\sfrac12(1{-}\psi^2)^2\ ,
\end{equation}
where $V_0$ is the value of~$V$ at the turning points.

\bigskip

\section{Magnetic and electric-magnetic Yang--Mills configurations on dS$_4$}

\noindent
Let us look at the gauge potential and field and compute its energy and action in terms of~$\psi$. 
Inserting (\ref{3.5}) into (\ref{3.2}), we obtain
\begin{equation}\label{3.6}
\Acal \=\sfrac12\, (1+\psi)\,e^a I_a
\qquad\textrm{and}\qquad
\Fcal \=\bigl(\sfrac12\,\dot\psi\,e^0\wedge e^a-\sfrac14\,(1{-}\psi^2)\,\ve^a_{bc}\,e^b\wedge e^c\bigr)I_a\ ,
\end{equation}
which yields the color-electric and -magnetic field components (in the cylinder metric)
\begin{equation}\label{EB}
E_a \= \Fcal_{0a} \= \sfrac12\,\dot\psi\,I_a \qquad\textrm{and}\qquad
B_a \= \sfrac12\ve_{abc}\Fcal_{bc} \= -\sfrac12\,(1{-}\psi^2)\,I_a\ .
\end{equation}
The electric and magnetic energy densities then become
\begin{equation}
\rho_e \= -\sfrac14\tr\,E_a E_a \= \sfrac34\,C(j)\,\dot\psi^2 \qquad\textrm{and}\qquad
\rho_m \= -\sfrac14\tr\,B_a B_a \= \sfrac34\,C(j)\,(1{-}\psi^2)^2\ ,
\end{equation}
respectively.
The energy of our Yang--Mills configuration (in the cylinder metric) computes to
\begin{equation} \label{energy1}
E_t \= \int_{S^3}\!e^1{\wedge}e^2{\wedge}e^3\;(\rho_e+\rho_m) 
\= \sfrac34\,C(j)\,\textrm{vol}(S^3)\,\bigl(\dot\psi^2+(1{-}\psi^2)^2\bigr)
\= 3\pi^2\,C(j)\,V_0\ ,
\end{equation}
where we have employed the energy relation~(\ref{3.10}) in the last step.
Remarkably, $E_t$ is constant and only given by the `double-well energy'~$V_0$.
It is important to note, however, that $E_t$ is conjugate to the time variable~$t$,
and so the energy conjugate to de Sitter time~$\tilde\tau$ (\ref{2.4}) is obtained as
\begin{equation} \label{energy2}
E_{\tilde\tau} \= \frac{\diff t}{\diff\tilde\tau}\,E_t \= \frac1R\,\frac{\diff t}{\diff\tau}\,E_t
\=\frac{1}{R\cosh\tau}\,E_t \= \frac{3\pi^2\,C(j)\,V_0}{R \cosh\tau}\ . 
\end{equation}
We see that this energy decays exponentially for early and late times.

In a similar fashion one can evaluate the action functional on the configuration (\ref{EB}).
Its value is independent of the metric chosen in the computation. For the cylinder~(\ref{2.8}),
for example, we get
\begin{equation}\label{action1}
\begin{aligned}
S &\= \sfrac18 \int_{{\cal I}\times S^3} \!\!\!\!\! e^0{\wedge}e^1{\wedge}e^2{\wedge}e^3\;
\tr(- 2\Fcal_{0a}\Fcal_{0a} + \Fcal_{ab}\Fcal_{ab}) 
\= \int_{\cal I}\!\diff t\ \textrm{vol}(S^3)\,(\rho_e-\rho_m) \\
&\= \sfrac{3}{2}\pi^2\,C(j) \int_{-\pi/2}^{\pi/2}\!\!\!\!\diff t\;\bigl(\dot\psi^2-(1{-}\psi^2)^2\bigr)
\= 3\pi^3\,C(j)\,V_0 
\ -\ 6\pi^2\,C(j)\int_{-\pi/2}^{\pi/2}\!\!\!\!\diff t\;V\bigl(\psi(t)\bigr)\ ,
\end{aligned}
\end{equation}
which takes a finite value for any solution $\psi(t)$ to~(\ref{3.7}).

For alternatively computing the action directly with the de Sitter metric~(\ref{2.4}), 
we introduce on orthonormal basis on dS$_4$,
\begin{equation}
\tilde{e}^0: = R\,\diff\tau , \ \tilde{e}^a: = R\cosh\tau\, e^a \ ,
\end{equation}
and expand
\begin{equation}\label{3.2tilde}
\Acal \= \Act_a\,\tilde{e}^a \qquad\textrm{and}\qquad
\Fcal \= \Fct_{0a}\,\tilde{e}^0\wedge \tilde{e}^a + \sfrac12 \Fct_{bc}\,\tilde{e}^b\wedge \tilde{e}^c
\end{equation}
so that
\begin{equation}
\Acal_a = R\cosh\tau\,\Act_a\ ,\quad
\Fcal_{bc} = R^2\cosh^2\!\tau\,\Fct_{bc}\ ,\quad
\Fcal_{0a} = \pa_t \Acal_a = R^2\cosh^2\!\tau\,\pa_{\tilde\tau}\Act_a\ .
\end{equation}
With these ingredients, we find that
\begin{equation}\label{action2}
S \=  \sfrac18 \int_{\textrm{dS$_4$}} \!\!\!\! 
\tilde{e}^0{\wedge}\tilde{e}^1{\wedge}\tilde{e}^2{\wedge}\tilde{e}^3\;
\tr(- 2\Fct_{0a}\Fct_{0a} + \Fct_{ab}\Fct_{ab})
\= \int_{\R}\!\diff\tau\ \textrm{vol}(S^3)\,\frac{\rho_e-\rho_m}{\cosh\tau}\ ,
\end{equation}
which agrees with (\ref{action1}) by virtue of~(\ref{2.6}) 
(note that $\int\diff\tau/\cosh\tau$ also produces the factor of~$\pi$).

\bigskip

\section{Explicit examples}

\noindent 
Finally, we analytically display the Yang--Mills configurations for the explicit solutions
(\ref{psiconst}) and~(\ref{3.9}).
The solutions $\psi(t)=\pm1$ correspond to the vacuum $\Fcal=0$ which is not interesting.
In contrast, $\psi(t)=0$ provides the nontrivial smooth configuration~\footnote{
We remark that that $e^aI_a = g^{-1}\diff g$, 
where $g(\th^a): S^3\to\,$SU(2) is a smooth map of degree (winding number) one.}
\begin{subequations}\label{4.2}
\begin{eqnarray}
\Acal \!&=&\! \sfrac12\,e^a\,I_a \=
\frac{1}{2R\cosh\tau}\,\tilde{e}^a\,I_a \ ,\\[4pt]
\Fcal \!&=&\! -\sfrac14\,\ve^a_{bc}\,e^b{\wedge}e^c\,I_a \=
-\frac{1}{4R^2\cosh^2\!\tau}\,\ve^a_{bc}\,\tilde{e}^b{\wedge}\tilde{e}^c\,I_a\ ,
\end{eqnarray}
\end{subequations}
hence (in the de Sitter metric)
\begin{equation}
\Et_a \= 0 \qquad\textrm{and}\qquad \Bt_a \= -\frac{I_a}{2\,R^2\cosh^2\!\tau}\ .
\end{equation}
This is a purely magnetic Yang--Mills field uniform on $S^3$ which varies with time~$\tau$
and decays exponentially for $\tau\to\pm\infty$.
According to (\ref{energy2}) with $V_0{=}\sfrac12$, the de Sitter energy of this configuration is finite,
\begin{equation}\label{4.3}
E_{\tilde\tau} \= \frac{3\pi^2\,C(j)}{2 R\cosh\tau}\ ,
\end{equation}
and the action is as well,
\begin{equation}\label{4.4}
S \= -\frac{3}{2}\pi^3\,C(j)\ ,
\end{equation}
gleaned from (\ref{action1}).
One may restore the gauge coupling in the denominator of (\ref{4.4}).

From the bounce (\ref{3.9}) we obtain a whole family of nonsingular Yang--Mills configurations,
\begin{subequations}\label{4.6}
\begin{eqnarray}
\Acal \!&=&\! \frac{\cos t}{2R}\,\biggl\{
1\ +\ \frac{\sqrt{2}}{\cosh\bigl(\sqrt{2}(t{-}t_0)\bigr)}\biggr\}\,\tilde{e}^a\,I_a\ ,\\[4pt]
\Fcal \!&=&\! -\frac{\cos^2\!t}{4R^2}\,\biggl\{
4\frac{\sinh\bigl(\sqrt{2}(t{-}t_0)\bigr)}{\cosh^2\bigl(\sqrt{2}(t{-}t_0)\bigr)}\,
\tilde{e}^0{\wedge}\tilde{e}^a\ +\ 
\Bigl(1 - \frac{2}{\cosh^2\bigl(\sqrt{2}(t{-}t_0)\bigr)}\Bigr)\,
\ve^a_{bc}\,\tilde{e}^b{\wedge}\tilde{e}^c \biggr\} I_a
\end{eqnarray}
\end{subequations}
depending on $t_0\in\R$, where $t=t(\tau)$ via (\ref{2.6}) is understood.
Clearly this family carries electric as well as magnetic fields. 
Since the bounce also has $V_0=\sfrac12$, the energy of this family coincides with that of
the above purely magnetic configuration, given by~(\ref{4.3}).
Its action, however, is different: From (\ref{action1}) we find
\begin{equation}
\frac{S}{C(j)} \= -\frac32\pi^3 + 12\pi^2\!\!\int_{-\pi/2}^{\pi/2}\!\!\!\!\diff t\
\frac{\sinh^2\bigl(\sqrt{2}(t{-}t_0)\bigr)}{\cosh^4\bigl(\sqrt{2}(t{-}t_0)\bigr)}
\= -\frac32\pi^3 + \sqrt{8}\pi^2 
\bigl( \tanh^3(\sfrac{\pi}{\sqrt{2}}{+}\delta) + \tanh^3(\sfrac{\pi}{\sqrt{2}}{-}\delta) \bigr)\ ,
\end{equation}
where $\delta=\sqrt{2}\,t_0\in\R$. Its numerical value varies between 5.52 (for $\delta{=}0$)
and -46.51 (for $\delta\to\pm\infty$).

In fact, for any choice of turning point and time, $V_0=V(\psi_0)$ and $\psi_0=\psi(t_0)$,
there is a unique solution $\psi(t)$ which gives rise to a smooth and $S^3$-homogeneous
Yang--Mills solution~$\Fcal$ with both color-electric and color-magnetic fields present.
All their energies and actions remain finite. 
As a final example, 
consider small oscillations about the vacuum $\psi=1$ in harmonic approximation,
\begin{equation}
\psi(t) \= 1+A\cos\bigl(2(t{-}t_0)\bigr) \qquad\Longrightarrow\qquad V_0 = -\sfrac12+2\,A^2\ .
\end{equation}
Neglecting terms of order $A^3$, it is easily calculated that in this case, independent of~$t_0$, 
\begin{equation}
\rho_e \= \rho_m \= \sfrac34\,C(j)\,A^2 \qquad\Longrightarrow\qquad
E_t = 6\pi^2\,C(j)\,A^2 \quad\textrm{and}\quad S=0\ .
\end{equation}

\bigskip

In summary, we have described a class of classical pure Yang--Mills configurations 
(without Higgs fields) on de Sitter space dS$_4$, which are spatially homogeneous
and decay for early and late times. Their energies and actions are all finite.
Therefore, the described gauge configurations can be important in a semiclassical analysis 
of the path integral for quantum Yang--Mills theory on dS$_4$.
These Yang--Mills solutions may help in understanding the dual superconducting mechanism 
of confinement on de Sitter space.

\bigskip

\noindent {\bf Acknowledgements}

\noindent 
This work was partially supported by the Deutsche Forschungsgemeinschaft grant LE~838/13 
and by the Heisenberg-Landau program.


\end{document}